\documentclass[a4paper]{article}
\usepackage{listings}
\usepackage{xcolor}
\usepackage{multirow}
\usepackage{jheppub}  
\usepackage{dcolumn}
\usepackage[english]{babel}
\usepackage[utf8x]{inputenc}
\usepackage[T1]{fontenc}
\usepackage{palatino}
\usepackage{slashed}
\usepackage{centernot}

\usepackage{amsmath}
\usepackage{afterpage}
\usepackage{graphicx, subcaption}
\usepackage[colorinlistoftodos]{todonotes}
\usepackage[colorlinks=true]{hyperref}
\usepackage{makeidx}

\newcommand{\be}{\begin{equation}}  
\newcommand{\ee}{\end{equation}}  
\newcommand{\bea}{\begin{eqnarray}}  
\newcommand{\eea}{\end{eqnarray}}  

\newcommand{\nn}{\nonumber}
\newcommand{\beqa}{\begin{eqnarray}}
\newcommand{\eeqa}{\end{eqnarray}}
\newcommand{\ud}{\mathrm{d}}
\newcommand{\uvec}{\boldsymbol}
\newcommand{\bd}[1]{ \mbox{\boldmath $#1$}}
\newcommand{\fsl}[1]{{\centernot{#1}}}
\newcommand{\shrink}[1]{\text{\scriptsize $#1$}}

\makeindex
\begin{document}

\vspace*{1.2cm}

\thispagestyle{empty}
\begin{center}
{\LARGE \bf Study of the potential transverse momentum and potential angular momentum
within the scalar diquark model}

\par\vspace*{7mm}\par

{

\bigskip

\large \bf David Arturo Amor-Quiroz$^1$ \\ Matthias Burkardt$^2$ \\William Focillon$^1$\\ C\'edric Lorc\'e$^1$}



\bigskip

{\large \bf  E-Mail: arturo.amor-quiroz@polytechnique.edu \\
burkardt@nmsu.edu \\
william.focillon@polytechnique.edu \\ 
cedric.lorce@polytechnique.edu
}

\bigskip

{$^1$CPHT, CNRS, Ecole Polytechnique, Institut Polytechnique de Paris, Route de Saclay, 91128 Palaiseau, France. \\
$^2$Department of Physics, New Mexico State, University, Las Cruces, New Mexico 88003, USA.}

\bigskip

{\it Presented at the Workshop of QCD and Forward Physics at the EIC, the LHC, and Cosmic Ray Physics in Guanajuato, Mexico, November 18-21 2019}


\vspace*{15mm}

{  \bf  Abstract }

\end{center}
\vspace*{1mm}

\begin{abstract}

We make use of a simple scalar diquark model to study the potential transverse momentum and potential angular momentum, defined as the difference between the Jaffe-Manohar and Ji notions of transverse momentum and orbital angular momentum, respectively. 
A non-vanishing potential angular momentum has been previously found in lattice calculations and is believed to appear due to the effects of initial/final state interactions between the spectator system and the struck quark in high energy scattering processes. Such re-scattering phenomena are similar in nature to those who are responsible for generating the Sivers shift. This motivates us to search for an estimate of the potential angular momentum in terms of the expectation value of the transverse momentum of the struck quark.
\end{abstract}
  
\section{Introduction}
\label{Intro}

One of the goals to be addressed by future experiments at the Electron Ion Collider (EIC) at the Brookhaven National Laboratory is to understand the origin of the spin of the proton \cite{Accardi:2012qut,Aidala:2020mzt}. From a theoretical point of view, achieving such an ambitious objective requires a proper decomposition of the nucleon total angular momentum (AM) into the orbital motion and intrinsic spin of its constituents. 
While many decompositions with different properties and physical interpretations are possible, the most common decompositions of AM are the Jaffe-Manohar (JM)~\cite{Jaffe:1989jz} and Ji~\cite{Ji:1996ek} decompositions.
They induce different notions of orbital angular momentum (OAM) that depend on how the interacting contributions are attributed to either quarks or gluons. 
Providing a broader insight on other kinds of decompositions is beyond the scope of the present document. Nevertheless, for a clear explanation of the differences and the physical interpretation of some standard decompositions, the reader is referred to the reviews by Leader and Lorc\'e~\cite{Leader:2013jra,leader2019corrigendum} and Wakamatsu~\cite{Wakamatsu:2014zza}.

The fundamental difference between Ji and JM decompositions of AM is that the former is related to the usual gauge covariant derivative $D_\mu$, while the latter is associated with the \textit{pure} gauge covariant derivative $D^{\mu}_{\text{pure}}$ defined as $D^{\mu}_{\text{pure}}=\partial^\mu-ig A^{\mu}_{\text{pure}}$. The notation refers to the \textit{Chen et al.}~\cite{Chen:2008ag} splitting of the (generally non-abelian) fields provided by~\cite{Hatta:2011ku,Lorce:2012ce}
\beqa
    A^\mu &=& A^{\mu}_{\text{pure}}+A^{\mu}_{\text{phys}} 
    \nn\\
    F^{\mu\nu}_{\text{pure}} &=&
    \partial^\mu A^{\nu}_{\text{pure}}
    -\partial^\nu A^{\mu}_{\text{pure}}
    -ig[A^{\mu}_{\text{pure}},A^{\nu}_{\text{pure}}]
    = 0 ~,
\eeqa
where gauge transformations act differently on the pure and physical fields.
The \textit{physical} part of the field ${A}^{\mu}_{\text{phys}}$ can be fixed by a condition similar to a gauge-fixing. We choose the light-front (LF) condition $A^{+}_{\text{phys}}=0$ since it appears to be convenient for simplifying some calculations, as the pure gauge derivative ${D}^{\mu}_{\text{pure}}$ coincides with the common partial derivative provided by $\partial^\mu$ in the corresponding LF gauge. We also use advanced boundary conditions and omit any label indicating such a choice.


Let us start with the potential AM, defined as the difference between the Ji and Jaffe-Manohar (JM) notions of OAM   $L^z_{\text{pot}}\equiv L^z_{\text{JM}}-L^z_{\text{Ji}}$ where~\cite{Wakamatsu:2010qj,Hatta:2011ku}
\beqa
&
L^z_{\text{Ji}}
\equiv 
\lim_{\Delta\to 0}\frac{\langle p',S'|\int \ud^4 r\,\delta(r^+)\,\overline\psi(r)\gamma^+[\uvec r_\perp\times i\uvec D_{\perp}]_z\psi(r)|p,S\rangle}{\langle P,S|P,S\rangle} ~,
& \label{DefJi}
\eeqa
is the Ji definition for the quark OAM inside a nucleon with initial (final) four-momentum $p^{(\prime)}$ and covariant spin $S^{(\prime)}$, while $P=\frac{p+p^\prime}{2}$ is the average 4-momentum. On the other hand, the JM notion of quark OAM in the LF gauge is provided by
\beqa
&
L^z_{\text{JM}}
\equiv
\lim_{\Delta\to 0}\frac{\langle p',S'|\int \ud^4 r\,\delta(r^+)\,\overline\psi(r)\gamma^+[\uvec r_\perp\times i\uvec D_{\text{pure},\perp}]_z\psi(r)|p,S\rangle}{\langle P,S|P,S\rangle}
~,
& \label{DefJM}
\eeqa
The computation of the matrix elements should first be considered
with $p' \ne p$, and the forward limit ${\Delta=p'-p\to 0}$ has to be taken at the end. 
We also consider initial and final nucleon states with the same rest-frame polarization $s=s^\prime = (\bd{s}_\perp , s_z)$ so that the covariant spin vector\footnote{ In light-front coordinates any 4-vector is given by 
$x^\mu=[x^+,x^-,\bd{x}_\perp]$ with ${x^{\pm}=\frac{1}{\sqrt{2}}(x^0 \pm x^3)}$.} is given by 
\beqa
S^\mu = [s_z P^+, -s_z P^- + \frac{\shrink{\bd{P}}_\perp}{P^+}\cdot (M\bd{s}_\perp + \bd{P}_\perp s_z), M\bd{s}_\perp + \bd{P}_\perp s_z]
\eeqa 
satisfying $P\cdot S=0$ and $S^2=-M^2$.

By means of Eqs. \eqref{DefJi} and \eqref{DefJM} it is easy to verify that the difference between the Ji and JM notions of the longitudinal component of the quark OAM in the LF gauge is proportional to the gauge fields.
The potential AM is consequently provided by~\cite{Wakamatsu:2010qj}
\beqa
&
L^z_{\text{pot}}
= 
\lim_{\Delta\to 0}\frac{\langle p',S'|\int \ud^4 r\,\delta(r^+)\,(-g)\overline\psi(r)\gamma^+[\uvec r_\perp\times\uvec A_{\text{phys},\perp}(r)]_z\psi(r)|p,S\rangle}{\langle P,S|P,S\rangle}\nn\\
&=\frac{-g\,\epsilon^{ij}_\perp}{2P^+}\left[-i\nabla^i_{\Delta_\perp}\langle p',S'|\overline\psi(0)\gamma^+A^j_{\text{phys},\perp}(0)\psi(0)|p,S\rangle\right]_{\Delta=0} . \quad
& \label{defLpot}
\eeqa
The potential AM is interpreted as the accumulated change in OAM experienced by the struck quark due to the color Lorentz forces as it leaves the target in high-energy scattering processes~\cite{Burkardt:2012sd}.

Similarly, the potential transverse momentum (TM) is defined as the difference between JM and Ji notions of the transverse momentum of a parton inside the nucleon~\cite{Wakamatsu:2010qj}
\beqa
&
\langle k^i_\perp\rangle_{\text{pot}} \equiv
\langle k^i_\perp \rangle_{\text{JM}} - \langle k^i_\perp \rangle_{\text{Ji}}
= 
\frac{-g}{2P^+}\,\langle P,S|\overline\psi(0)\gamma^+A^i_{\text{phys},\perp}(0)\psi(0)|P,S\rangle
 \,.
& \label{defkpot}
\eeqa
The potential TM can be related to the Sivers shift (as justified with more detail in Section \ref{Discussion}), which is the non-vanishing average transverse momentum of individual partons orthogonal to the nucleon transverse spin resulting from the Sivers mechanism.

Using perturbation theory, explicit calculations found that $L_{\text{pot}}$ vanishes at tree level~\cite{Ji:2015sio}. The aim of the present document is to check within a model calculation wether the difference between Ji and JM decompositions of OAM appears at two-loop level and to compare its magnitude with that of $\bd{k}_{\text{pot}}$~\cite{Burkardt:2012sd}. 

A non-vanishing potential AM is supported by recent lattice calculations which demonstrate that it can be clearly resolved and furthermore, that the JM OAM appears to be ``significantly enhanced'' when compared with Ji OAM~\cite{Engelhardt:2017miy,Engelhardt:2019lyy}. Additionally, the renormalization scale dependence of $L_{\text{pot}}$ was recently studied in lattice QCD by Hatta and Yao~\cite{Hatta:2019csj} with results that seem to be compatible with those previously found in~\cite{Ji:2015sio}. For a recent discussion of the potential AM in the Landau problem, see~\cite{Wakamatsu:2017isl}.

Both the potential TM and the potential AM are gauge invariant quantities that can in principle be experimentally observed.
%
%
%
%
%
The reason for comparing their magnitudes is motivated by Burkardt's proposal of defining a lensing effect due to soft gluon rescattering in deep-inelastic  and  other  high-energy scatterings. Originally, such effect was described by factorizing the Sivers function $f^{\perp}_{1T}$ into a distortion effect in position space (described by the distortion GPD $E$)  times a lensing function $\mathcal{I}(\bd{b}_\perp)$ that accounts for the effect of attractive initial/final state interactions (ISI/FSI) in the impact parameter space denoted by $\bd{b}_\perp$. Such interactions are responsible for providing both TM to the outgoing quark, as well as exerting a torque on it as it leaves the target~\cite{Burkardt:2002ks,Burkardt:2003uw}. 

Even if the notion of a lensing function is intuitive and is supported by some model calculations~\cite{Burkardt:2003uw,Burkardt:2003je,Bacchetta:2011gx,Gamberg:2009uk}, we emphasize that the possibility of factorizing $\mathcal{I}(\bd{b}_\perp)$ is model dependent~\cite{Pasquini:2019evu}.
We argue, nevertheless, that a non-vanishing Sivers function is both a direct probe of orbital motion of the partons inside the nucleus and a necessary condition for a non-vanishing potential AM, as both mechanisms have the same physical foundations in ISI/FSI.\\

In the present document, we compute both the potential TM and potential AM in the framework of a simple scalar diquark model (SDM) of the nucleon at $\mathcal{O}(\lambda^2 e_q e_s)$ in perturbation theory. The model is  based on the assumption that the nucleon splits into a quark and a scalar diquark structure, which are regarded as elementary fields of the theory~\cite{Brodsky:2002cx}. The lagrangian for the SDM is
\begin{align}
\mathcal{L}_{\text{SDM}} 
&=
-\frac{1}{4}\,F^{\mu\nu}F_{\mu\nu}+
\bar{\Psi}_{N} \left( i\fsl{\partial}-M \right) \Psi_{N}
+
\bar{\psi} \left( i\fsl{\partial}-m_q \right) \psi
+
\partial_\mu\phi \partial^\mu\phi^* 
- 
m^2_s \phi^* \phi
\nn\\
& +
\lambda  \left( \bar{\psi}\Psi_{N}\phi^* + \bar{\Psi}_{N}\psi\phi \right) 
-
e_q \bar{\psi} \fsl{A} \psi
-
i e_s \big( \phi^* \overset{\leftrightarrow}{\partial^\mu} \phi \big) A_\mu
+
e^2_s A_\mu A^\mu \phi^* \phi ~.
\quad
\label{Lagrangian}
\end{align}
In this expression, $\psi$ represents the quark field with mass $m_q$; $\phi$ denotes the charged scalar diquark field with mass $m_s$; $\Psi_N$ is the neutral nucleon field with mass $M$; and $A$ are abelian gauge fields that could represent either photons or gluons, as any non-abelian effects would appear at three-loop order.
The stability condition for the target nucleon is $M<m_q+m_s$.
Furthermore, the photon-quark and photon-diquark couplings are given by $e_q$ and $e_s$ respectively, while $\lambda$ is the coupling constant of the point-like scalar quark-nucleon-diquark vertex. 

The target field considered in the present model has no charge, emulating either a neutron in QED or the fact that in QCD a hadron is color neutral. Such condition simplifies the calculations as no gauge fields can couple to the target. Moreover, it implies that the photon-quark and photon-diquark coupling constants are equal but with opposite sign, for instance, $e_q=-e_s$ for QED.
For the corresponding QCD generalization of the results in the following sections, it suffices to do the replacement $e^2_q \rightarrow 4\pi C_F \alpha_s$~\cite{Brodsky:2002cx}.

The SDM has the feature of maintaining explicit Lorentz covariance and provides analytic results that have broadly been explored in the literature. Because of this, the SDM presents a good framework for providing an estimate of the magnitude of potential AM with respect to the magnitude of potential TM. The effect of introducing a vector diquark on said observables is beyond the interest of the present document. \\

The manuscript is organized as follows. In Sec.~\ref{Discussion} we define and compute the  potential TM of an unpolarized parton in a transversely polarized nucleon.
Thereafter we report the potential AM in Section \ref{PotOAM} for a parton in a longitudinally polarized nucleon and provide an estimate of its magnitude with respect to the potential TM. Finally, we summarize our results in Sec.~\ref{Conclusions}. \\

\section{The transverse momentum of unpolarized quarks inside a transversely polarized nucleon}
\label{Discussion}

A non-zero average transverse momentum of a parton inside a nucleon (and eventually a non-vanishing potential TM) implies the existence of a privileged direction that breaks the cylindrical symmetry induced by the axis of propagation of the nucleon. Such phenomenon can only arise due to the presence of spin correlations in the plane transverse to said propagation.

The mathematical object of interest in the transverse plane is
the transverse momentum dependent correlation function, which is defined as
\beqa
\Phi^{[\gamma^+]}(P,x,\bd{k}_\perp,S)
= \frac{1}{2}\int\frac{\ud z^- \ud^2 z_\perp}{(2\pi)^3}e^{ik\cdot z} \langle P,S | \bar{\psi}(-\tfrac{z}{2})\gamma^+\mathcal{W}(-\tfrac{z}{2},\tfrac{z}{2})\psi(\tfrac{z}{2}) | P,S \rangle \Big|_{z^+=0} ~,
\label{TMDcorrelator}
\eeqa
where $\bd{k}_\perp$ is the quark transverse momentum inside a transversely polarized target with average 4-momentum $P$ and spin vector $S$. The parameter $x$ is the fraction of the total nucleon momentum carried by such quark along the LF direction. Furthermore, the LF gauge with advanced boundary conditions simplifies the Wilson line $\mathcal{W}(-\tfrac{z}{2},\tfrac{z}{2})\mapsto1$ that would otherwise appear explicitly in expression \eqref{TMDcorrelator} to ensure the gauge invariance.

The correlation function $\Phi^{[\gamma^+]}$ can be parametrized in terms of transverse-momentum
distributions (TMDs).
They can be interpreted in temrs of three-dimensional densities in momentum space. For the leading-twist vector operator we have for $\uvec P_\perp=\uvec 0_\perp$
\beqa
\Phi^{[\gamma^+]}(P,x,\bd{k}_\perp,S)
&=& 
f_1(x,\uvec{k}^2_\perp)
-
\frac{\epsilon^{ij}_\perp k^i_\perp S^j_\perp }{M}
f^{\perp }_{1T}(x,\uvec{k}^2_\perp).
\eeqa
We will focus only on the Sivers TMD ${f^{\perp }_{1T}}$ which corresponds to a correlation of the type $\bd{S}\cdot(\bd{P}\times\bd{k})$~\cite{Sivers:1989cc}. The Sivers function describes the left-right distortion in the distribution of partons known as the Sivers effect.
This effect was proposed by D. Sivers in 1990 as a way to explain large left-right asymmetries observed in pion-nucleus collisions~\cite{Antille:1980th}. This effect was thought to vanish due to time-reversal symmetry, but nowadays we know that processes that are odd under naive time reversal (T-odd) transformations allow the existence of mechanisms such as Sivers' \cite{Collins:2002kn}. 
The Sivers asymmetry has indeed been experimentally observed in both SIDIS~\cite{Airapetian:2004tw,Alexakhin:2005iw,Alekseev:2010rw} and Drell-Yan~\cite{Aghasyan:2017jop} processes.

In the following subsections we will discuss and try to clarify the relation between the Sivers function, the different notions of transverse momentum and the potential TM.

\subsection{JM notion of transverse momentum}
\label{JM-TransvMomentum}

The Sivers mechanism gives rise to a non-zero average parton transverse momentum (TM) of the type $a$ inside the nucleon by means of the following identity~\cite{Meissner:2007rx}
\beqa
 \langle k^{(a)i}_{\perp} \rangle_{\text{JM}} 
 =
 - \int \ud x  \int \ud^2 k_{\perp} \,
 {k}^{i}_{\perp} 
 \frac{\epsilon^{jk}_\perp{k}^{j}_{\perp} {S}^{k}_{\perp}}{M}
 f^{\perp a}_{1T}(x,\bd{k}^2_\perp) ~.
 \label{IdentitySiversk}
\eeqa
Due to the T-odd nature of the Sivers function, this notion of transverse momentum exactly coincides with the JM notion, as the latter refers to a non-local notion of the covariant derivative and can in principle contain a mixed symmetry under naive time reversal. The possible T-even contribution corresponding to the unpolarized TMD $f_1$ vanishes due to time reversal symmetry.

Moreover, total TM is conserved as long as the contributions from the partons sum up to zero in what is known as Burkardt sum rule, irrespective of the decomposition used~\cite{Burkardt:2004ur}. 
In the case of the SDM it takes the form     
${ \sum_{a=q,s} \langle \bd{k}^{(a)}_\perp \rangle  = \bd{0}_{\perp} }$ where $q$ and $s$ denote the quark and the scalar diquark, respectively. By using the expression for the quark Sivers function within the SDM provided in~\cite{Ji:2002aa}
\beqa 
f^{\perp q}_{1T}(x,\bd{k}^2_\perp)
&=&
\frac{\lambda^2 e_q e_s (1-x)}{4(2\pi)^4}
\frac{M(m_q + xM)}{(\bd{k}^{2}_{\perp}+\Lambda^2_q (x))}
\ln \left(
\frac{\bd{k}^{2}_{\perp}+\Lambda^2_q (x)}{\Lambda^2_q (x)}
\right)
\nn\\
\Lambda^2_q(x)
&=&
xm^2_s+(1-x)m^2_q-x(1-x)M^2
\label{JiGoekeSivers}
\eeqa
and by virtue of Eq.~\eqref{IdentitySiversk}, Goeke \textit{et al.} found that the Burkardt sum rule is fulfilled for the JM notion of transverse momentum~\cite{Goeke:2006ef}.

From the integration of Eq.~\eqref{IdentitySiversk} using the quark Sivers function in~\eqref{JiGoekeSivers} and dimensional regularization, we obtained that the leading divergent piece of the JM notion of TM is provided by
\beqa
\langle k^{(q)i}_{\perp} \rangle_{\text{JM}} 
= -
\frac{1}{6}
\left( \frac{\lambda}{4 \pi\epsilon} \right)^2
\frac{\epsilon^{ij}_{\perp} s^{j}_\perp }{ (4\pi)^2}(3m_q + M)\pi e_s e_q + \mathcal{O}\left(\frac{1}{\epsilon}\right)~. 
& \label{k_average}
\eeqa
In this equation, $s_\perp$ denotes the transverse polarization of the target. The factor $4\pi\epsilon$ comes from the dimensional regularization for the integral over TM, meaning that only the leading divergence is retained. The calculation requires some care as the expression in Eq. \eqref{JiGoekeSivers} results from a previous integral
\beqa
\int \frac{\ud^2 l_\perp}{ (2\pi)^2}
\frac{l^{j}_\perp}
{\bd{l}^{2}_\perp[(\bd{k}^{2}_{\perp}+\bd{l}^{2}_{\perp})+\Lambda^2_q (x)]}
= -
\frac{k^{j}_{\perp}}{4\pi \bd{k}^{2}_{\perp}}
\ln \left(
\frac{\bd{k}^{2}_{\perp}+\Lambda^2_q (x)}{\Lambda^2_q (x)}
\right)
\eeqa
which is finite in $D=2$, that also leads to contributions of $\mathcal{O}(\epsilon^{-2})$ when making the replacement $D=2\rightarrow 2-2\epsilon$ in the integrals over the momenta $\bd{k}^{2}_{\perp}$ and $\bd{l}^{2}_{\perp}$ at the same time.

\subsection{Potential transverse momentum}
\label{PotMomentum}

It is due to the local description of the full covariant derivative that Ji notion of partons TM is even under naive time-reversal transformations (T-even) and therefore it has to vanish to all orders in perturbation theory.
This means that the potential TM is equal to the JM notion of TM, i.e., $\langle\bd{k}_\perp\rangle_{\text{pot}} = \langle \bd{k}_\perp \rangle_{\text{JM}}$,
where the latter can be obtained from Eq.~\eqref{IdentitySiversk}. The potential TM is therefore a pure naive T-odd function.

We corroborated by means of an explicit calculation that the naive T-even contributions to the potential TM do vanish up to $\mathcal{O}(\lambda^2 e_q e_s)$.
The non-vanishing difference between the Ji and JM notions of TM appears only when one photon is attached to the spectator system, as a naive T-odd contribution comes from the gauge-link at the LF infinity.
This supports the interpretation of such a difference as originating from ISI/FSI between the spectator system and the struck quark. 

The expression we found for $\langle\bd{k}_\perp\rangle_{\text{pot}}$ coincide exactly with the result displayed in Eq.~\eqref{k_average} in agreement with the previous statement that
$ {\langle\bd{k}_\perp\rangle_{\text{pot}} = \langle \bd{k}_\perp \rangle_{\text{JM}} }$. 
Furthermore, Burkardt sum rule for the JM notion of TM is confirmed for the two-loop calculation by using the equivalent expression
\beqa
 \sum_{a=q,s} \langle\bd{k}^{(a)}_\perp\rangle_{\text{pot}}  = \bd{0}_{\perp}.
\eeqa
 For the sake of completeness, we shall mention that Burkardt sum rule is trivially fulfilled for the Ji decomposition as ${ \langle \bd{k}^{(q)}_\perp \rangle_{\text{Ji}} 
= \langle \bd{k}^{(s)}_\perp \rangle_{\text{Ji}} 
= \bd{0}_{\perp} } $. 
This is supported by the physical intuition given that Ji decomposition excludes the effects of ISI/FSI that can provide a spin asymmetry. 

\section{Quark potential AM in a longitudinally polarized nucleon}
\label{PotOAM}

Using the LF gauge it has also been computed that the Ji and JM notions of OAM coincide ($L^z_{\text{Ji}} =L^z_{\text{JM}} $)  in the absence of gauge fields within the SDM~\cite{Burkardt:2008ua}. Such result was later confirmed for the same model in the impact parameter space~\cite{Lorce:2017wkb}. Furthermore, a vanishing difference between both decompositions ($L^z_{\text{pot}}=0$) was proven directly for QED~\cite{Ji:2015sio}.

As a consequence, the potential AM vanishes at one-loop order and any difference between Ji and JM decompositions can only be observed at higher order. At tree level we are able to provide analytical expressions for the OAM distributions of the partons as
\beqa
   L^{q,z}_{\text{Ji}}(x) 
   &=& 
   \frac{ \lambda^2 }{4\pi}(1-x)^2 
   \int \frac{\ud^2 k_{\perp}}{(2\pi)^2}    
   \frac{ \bd{k}^2_\perp }{ ( \bd{k}^2_\perp + \Lambda^2_q(x) )^2 } \nn\\
   L^{s,z}_{\text{Ji}}(x) 
   &=&  
   \frac{x}{1-x} \,L^{q,z}_{\text{Ji}}(x)  ~,
   \label{1-loopOAM}
\eeqa
where $\Lambda^2_q(x)$ was defined in Eq.~\eqref{JiGoekeSivers} and $x\equiv x_q$ is the fraction of the nucleon 4-momentum carried by the quark in the LF direction. By momentum conservation, we can write for the diquark $x_s \equiv 1-x$.

Furthermore, at the same order in perturbation theory we can make use of the expression for the longitudinal helicity parton distribution function $g^q_{1L} (x)$ in~\cite{Meissner:2007rx}, which accounts for the contribution to the total AM coming from the quark spin projection along the propagation axis
\beqa
g^q_{1L} (x)
=
\frac{\lambda^2}{4\pi}(1-x)
\int \frac{\ud^2 k_{\perp}}{(2\pi)^2}  
\frac{(m_q+xM)^2 - \bd{k}^2_\perp }{ ( \bd{k}^2_\perp + \Lambda^2_q )^2 } ~.
\label{1-loopg}
\eeqa

From the expressions in Eq.~\eqref{1-loopOAM} and \eqref{1-loopg} it is possible to check that Ji sum rule~\cite{Ji:1996ek} is also fulfilled at this order within the SDM
\beqa
J = \int \ud x 
\left[ L^{q,z}_{\text{Ji}}(x) + L^{s,z}_{\text{Ji}}(x) + \frac{1}{2}\,g^q_{1L} (x) \right]
=\frac{1}{2} ~.
\eeqa\\
A first two-loop calculation of the potential AM as defined in Eq. \eqref{defLpot} will soon be reported in another publication.

\section{Summary and Outlook}
\label{Conclusions}
 
The potential transverse momentum was computed within the scalar diquark model as an alternative way of obtaining the difference between Ji and JM average transverse momenta, which turns out to be non-zero in perturbation theory at $\mathcal{O}(\lambda^2 e_q e_s)$.

\vspace*{-0.05cm}
 
In order to generate the Sivers effect the impact  parameter  distribution  of  unpolarized quarks in a transversely polarized target has to be distorted (non-zero $E$) and  the  fragmenting  struck  quark has to experience either initial or final state interactions.
A non-vanishing Sivers function therefore suggests a difference between Ji and JM decompositions to appear at best at two-loop order to include initial or final state interactions between the struck quark and the spectator system.

For the moment, only one-loop calculations were carried out in the literature for Ji and JM decompositions for OAM, where they coincide as expected from the lack of initial or final state interactions.
At this order it was also explicitly verified that Ji sum rule is satisfied. Further results on the two-loops calculation of the potential AM will be published in a future work.


\section*{Acknowledgements}

The author D. A. Amor-Quiroz acknowledges financial support from Consejo Nacional de Ciencia y Tecnolog{\'\i}a (CONACyT) postdoctoral
fellowship number 711226 (CVU 449539) and from Ecole Polytechnique. 
C. Lorc\'e by the Agence Nationale de la Recherche under the Projects No. ANR-18-ERC1-0002 and No. ANR-16-CE31-0019. M. Burkardt was supported by the DOE under grant number DE-FG03-95ER40965 and is grateful for the hospitality of Ecole Polytechnique.


\printindex

\end{document}